\documentclass[twocolumn]{aastex62}

\usepackage{amssymb}
\usepackage{amsmath}
\usepackage[]{graphicx}
\usepackage{enumerate}
\usepackage{subeqnarray}
\usepackage{cases}
\usepackage{mathrsfs,amssymb}

\usepackage{color} 
\usepackage{soul} 
\usepackage{xspace}

\newcommand{\Fermi}{\textit{Fermi}\xspace}

\tightenlines

\begin{document}

\title{ Gamma-Ray Emission from Molecular Clouds Generated by Penetrating Cosmic Rays}


\author{V.~A.~Dogiel}
\affiliation{I.E.Tamm Theoretical Physics Division of P.N.Lebedev Institute of Physics, Leninskii pr. 53, 119991 Moscow,
Russia}

\author{D.~O.~Chernyshov}
\affiliation{I.E.Tamm Theoretical Physics Division of P.N.Lebedev Institute of Physics, Leninskii pr. 53, 119991 Moscow,
Russia}
\affiliation{Moscow Institute of Physics and Technology (State University), 9, Institutsky lane, Dolgoprudny, 141707, Russia}

\author{A.~V.~Ivlev}
\affiliation{Max-Planck-Institut f\"{u}r Extraterrestrische Physik, Giessenbachstr. 1, D-85748 Garching, Germany}

\author{D.~V.~Malyshev}
\affiliation{Erlangen Centre for Astroparticle Physics, Erwin-Rommel-Str. 1, D-91058 Erlangen, Germany}

\author{A.~W.~Strong}
\affiliation{Max-Planck-Institut f\"{u}r Extraterrestrische Physik, Giessenbachstr. 1, D-85748 Garching, Germany}

\author{K.~S.~Cheng}
\affiliation{Department of Physics, University of Hong Kong, Pokfulam Road, Hong Kong, China}

\correspondingauthor{V.~A.~Dogiel}
\email{dogiel@td.lpi.ru}

\correspondingauthor{A.~V.~Ivlev}
\email{ivlev@mpe.mpg.de}


\begin{abstract}
We analyze the processes governing cosmic-ray (CR) penetration into molecular clouds and the resulting generation of
gamma-ray emission. The density of CRs inside a cloud is depleted at lower energies due to the self-excited MHD turbulence.
The depletion depends on the effective gas column density (``size'') of the cloud. We consider two different environments
where the depletion effect is expected to be observed. For the Central Molecular Zone, the expected range of CR energy
depletion is $E\lesssim 10$~GeV, leading to the depletion of gamma-ray flux below $E_\gamma\approx 2$~GeV. This effect can
be important for the interpretation of the GeV gamma-ray excess in the Galactic Center, which has been revealed from the
standard model of CR propagation (assuming the CR spectrum inside a cloud to be equal to the interstellar spectrum).
Furthermore, recent observations of some local molecular clouds suggest the depletion of the gamma-ray emission, indicating
possible self-modulation of the penetrating low-energy CRs.
\end{abstract}

\keywords{ISM: clouds --- cosmic rays -- gamma rays --- radiation mechanisms: non-thermal --- MHD turbulence --- scattering}

\section{Introduction}

The  diffuse gamma-ray emission observed in the Galaxy is an ubiquitous phenomenon, which is described well by standard
models of cosmic-ray (CR) interactions with the interstellar gas and photons \citep[see e.g.][as well as \citealp{strong04},
\citealp{acker12} and \citealp{acero16}]{ber90}. In the  Galactic plane gamma rays are mainly generated by interactions of
CR protons and nuclei with the interstellar gas;  bremsstrahlung and inverse Compton, produced by CR electrons, also
contribute to the emission. The nuclear collisions create $\pi^0$-mesons, which immediately decay into gamma-ray photons.
The rest energy of a $\pi^0$-meson is 135 MeV, which means that the kinetic energy of CR protons in the laboratory system
should be higher than $\approx280$~MeV.

Due to the threshold in the $\pi^0$ production, a depletion in the diffuse gamma-ray spectrum at $E_\gamma\lesssim100$~MeV
should be observed. Above this energy, the diffuse flux from the Galactic plane is mainly due to proton-proton (pp)
collisions. However, observations with the {\it Fermi} Large Area Telescope (LAT) in the direction of the Inner Galaxy
revealed a GeV excess, i.e., the flux above predictions of the standard CR propagation and interactions model, with a peak
in the spectrum around a few GeV \citep[see][]{good09,Vitale09,hooper11,abad12,hooper13, gordon13,
cal15,Zhou15,ajello16,Daylan16}. \citet{tim16} pointed out that the value of excess depends on the assumed spectrum of CRs
in the Galactic Center (GC).

A possible interpretation of the GeV excess is the  annihilation of DM around the GC \citep[see, e.g.,][]{hooper11,gher15,tim15}.
This interpretation is based on the fact that the hypothetical DM distribution in the Galaxy has a spherically symmetric
density distribution with a  peak at the GC \citep[see e.g.][for a general review see~\citealp{berg12}]{ber92,navar96}.
Millisecond pulsars were also suggested as sources of the excess from the GC \citep[see
e.g.][]{gnedin,gordon13,arca17,bartels,hooper18}, although CR interactions with the gas and photons cannot be excluded
either \citep[see e.g.][]{yang4, kwa}.


Recently, a self-consistent model of CR modulation in dense molecular clouds was proposed by \citet{ivlev18} (denoted below
as Paper I). While the inner structure of molecular clouds could be quite complex, in that paper we considered a simplified
model of a cloud, consisting of a core with a high gas density and a diffuse surrounding envelope. The core was assumed to
absorb completely the penetrating CR flux. It was shown that the modulation of the flux occurs due to CR scattering on
self-generated turbulence, excited by propagating CRs in the diffuse envelope. In the present paper we generalize the theory
of Paper I and investigate whether this can explain gamma-ray features observed in the GC and local molecular clouds. In
Section~\ref{fermi} we discuss the available observations of gamma-ray emission from GC and local molecular clouds, and
briefly analyze the processes that can affect the emission due to CR protons. In Section~\ref{self} we summarize important
results of the model of Paper I, which are necessary to calculate the expected depletion of the gamma-ray spectrum, and
investigate modifications introduced due to the finite cloud size and heavy CR species. In Sections~\ref{gamma_CMZ} and
\ref{gamma_local} we present the results of our calculations for the Central Molecular Zone (CMZ) and local clouds,
respectively, pointing out the expected features of the gamma-ray emission. Finally, in Section~\ref{discussion} we
summarize our major findings.

\section{Diffuse gamma-ray emission}
\label{fermi}

The CMZ is a region with some of the most dense molecular clouds. More than 90\% of its mass is concentrated in very dense
clouds (cores) of the average density over $10^4$ cm$^{-3}$ on the scale about several pc \citep[see e.g.][]{mezger,ao16}.
Compared to the clouds in other parts of the Galactic Disk, these clouds show unusually shallow density gradients
\citep[see][]{kauff}. On the other hand, the CMZ volume is mostly filled with a diffuse gas (envelope) surrounding the
cores, with the density of $\la 50$ cm$^{-3}$ \citep[see][]{oka05,petit,riquel}.

Thus, in the context of interactions with Galactic CRs, the CMZ can be treated as a very large cloud of the characteristic
size of $\sim 100$~pc \citep[see][]{katia}. The gas mass is concentrated in a single dense core with the line-of-sight
column density $N_{\rm H_2}\sim 3\times10^{23}$~cm$^{-2}$, surrounded by a diffuse envelope with the gas density of less
than $50$~cm$^{-3}$. At a distance of 8.5 kpc to the GC, the core has the angular size of $\sim 40'$, i.e., comparable to
the spatial resolution of the \Fermi LAT at $E_\gamma=1$~GeV. Therefore, it is a challenging task to separate the emission
generated by CRs in molecular clouds of the CMZ from other gamma-ray sources, including Sgr A$^*$.

The CMZ region contains several GeV point sources of unknown origin \citep[see the \Fermi source catalog][]{acero15}. For
example, \citet{malyshev15} concluded that the spectrum of one of these sources in the direction of Sgr A$^*$ does not show
a turnover at $E_\gamma\sim100$~MeV energy, expected for gamma rays originating from $\pi^0$-decay. One of the explanation
is that this emission is produced by high-energy electrons emitted by the central source \citep[see e.g.][and references
therein]{malyshev15,chern17}. Observations with the HESS telescope suggest that Sgr A$^*$ is a source of PeV CRs
\citep[see][]{abram16}.

Also, CRs can be locally produced in the CMZ when a supernova shock interacts with a nearby cloud, generating a gamma-ray
flux from the cloud. Several such sources were found by {\it Fermi}-LAT \citep[see e.g.][]{uchiyama}. This component of the
CMZ gamma-ray emission correlates with the star formation regions \citep[][]{tim16} and may also contribute to the total
flux. Moreover, CRs of relatively low energies (up to hundreds MeV or several GeV) can be directly accelerated inside
molecular clouds, e.g., in the inner regions of collapsing molecular clouds or circumstellar discs
\citep[][]{padovani16,padovani18}, but their contribution into the total CMZ flux still needs to be analyzed.

Another source of gamma-ray emission from the GC region may be the Fermi bubbles \citep[see][]{2010ApJ...724.1044S,
acker14}. In this case, the emission could be produced by CR protons which are generated in the star-forming regions of CMZ
and transported into the halo by a strong wind \citep[see][]{crock11,crock11b}. Alternatively, the emission could be due to
CRs which are accelerated at the Bubble edges by a shock resulting from stellar accretion onto the central black hole in the
past \citep[see][for the review of Bubble models, see also \citealt{yang18}]{cheng11}. The presence of the wind is detected
by the observation of UV absorption lines \citep[see][and references therein]{2018ApJ...860...98K} independently of the
model of the Fermi bubbles. This wind may modulate the CR spectrum in the inner Galaxy.


In addition to CMZ observations, the possible modulation of CR spectra at lower energies may be deduced from observations of
local molecular clouds, with the advantage that gamma-ray emission produced in such objects is measured directly
\citep[see][]{yang1,ner17,remy17}. However, we should take into account a substantial difference between the physical
parameters of the CMZ and of local molecular clouds, in particular the fact that the gas column density of local clouds does
not exceed $10^{21}-10^{22}$ cm$^{-2}$. Therefore, special conditions are needed for the CR modulation in the latter case,
as discussed in Section~\ref{gamma_local}.

We remind the reader that the goal of this paper is to focus specifically on the effect of self-modulation of CRs, occurring
upon their penetration into molecular clouds, on the resulting gamma-ray emission. As we pointed out above, this effect is
one of the many other effects (such as the local sources of CRs and the wind) that influence the resulting CR spectrum. A
comparison of the importance of the different effects that shape the spectrum of CRs near the GC is postponed to a future
work. At the moment, the self-modulation effect is not included in the CR propagation codes, such as GALPROP
\citep[][]{GALPROP}.


\subsection{Analysis of gamma-ray emission in the direction of CMZ and above}

For gamma-ray emission produced due to the interaction of CRs with molecular clouds of the CMZ, one would also expect to
observe the spectral variations in the direction perpendicular to the Galactic Plane. In particular, \cite{boer2} claimed
that there exists a change in the spectrum of the Galactic gamma-ray emission as a function of the latitude. They assumed
that the gamma-ray emission produced by CR protons consists of two components. One is generated inside molecular clouds, by
``modulated'' CRs with the energy spectrum which is harder than that in the interstellar medium; the other is generated
outside the clouds, with the CR spectrum similar to that near the Earth. According to their model, the ``excess'' emission
is seen directly from the CMZ ($-1.5^\circ\leq\ell\leq +2^\circ$ and $-0.5^\circ \leq b\leq +0.5^\circ$, where a significant
fraction of molecular hydrogen is concentrated), with the energy flux of $\approx 2\times10^{-6}$~GeV~s$^{-1}$~cm$^{-2}$ at
$E_\gamma\approx 2$~GeV. In their analysis they have not found any evidence for the GeV excess.

\begin{figure}[h]
\centering
\includegraphics[width=0.45\textwidth]{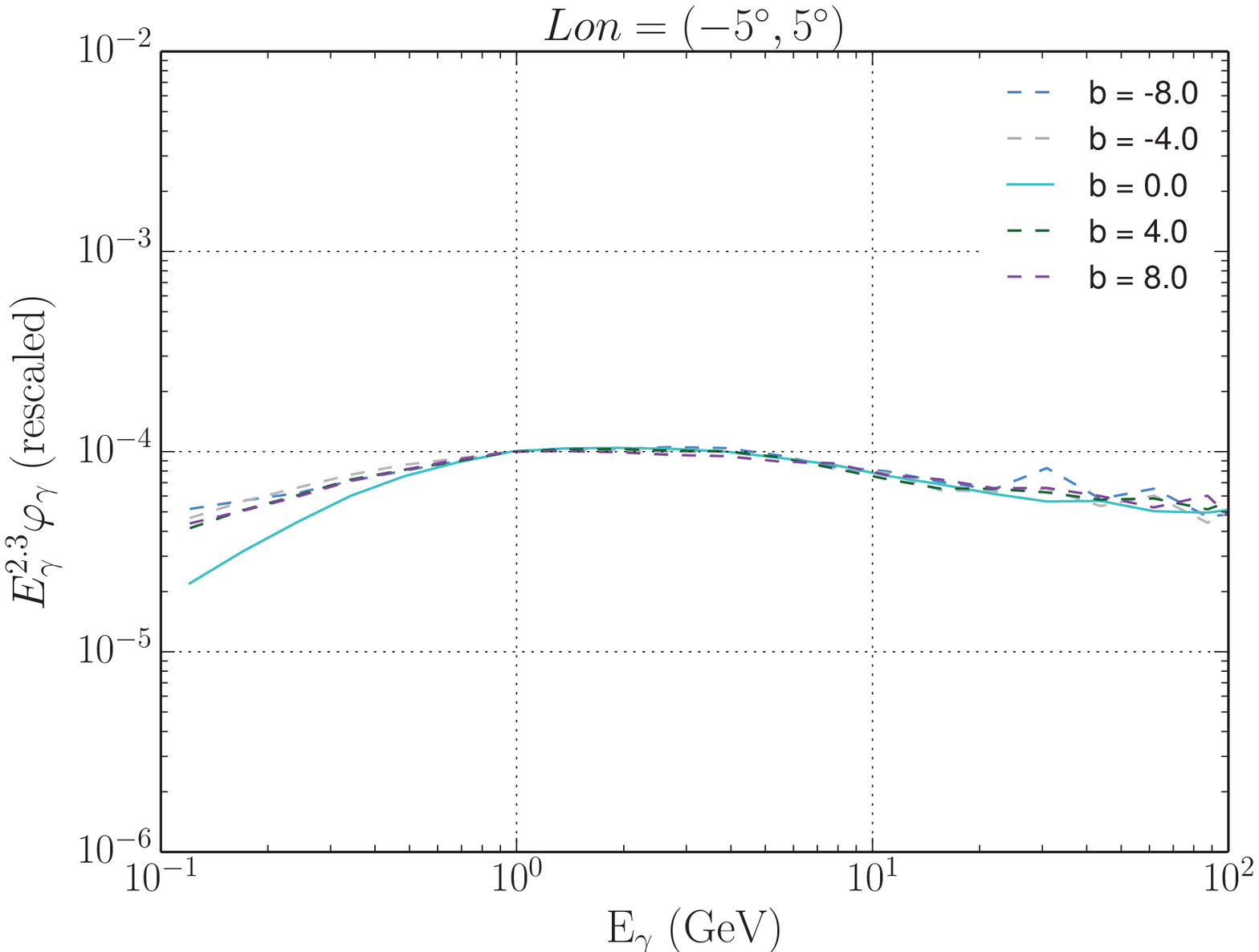}
\includegraphics[width=0.45\textwidth]{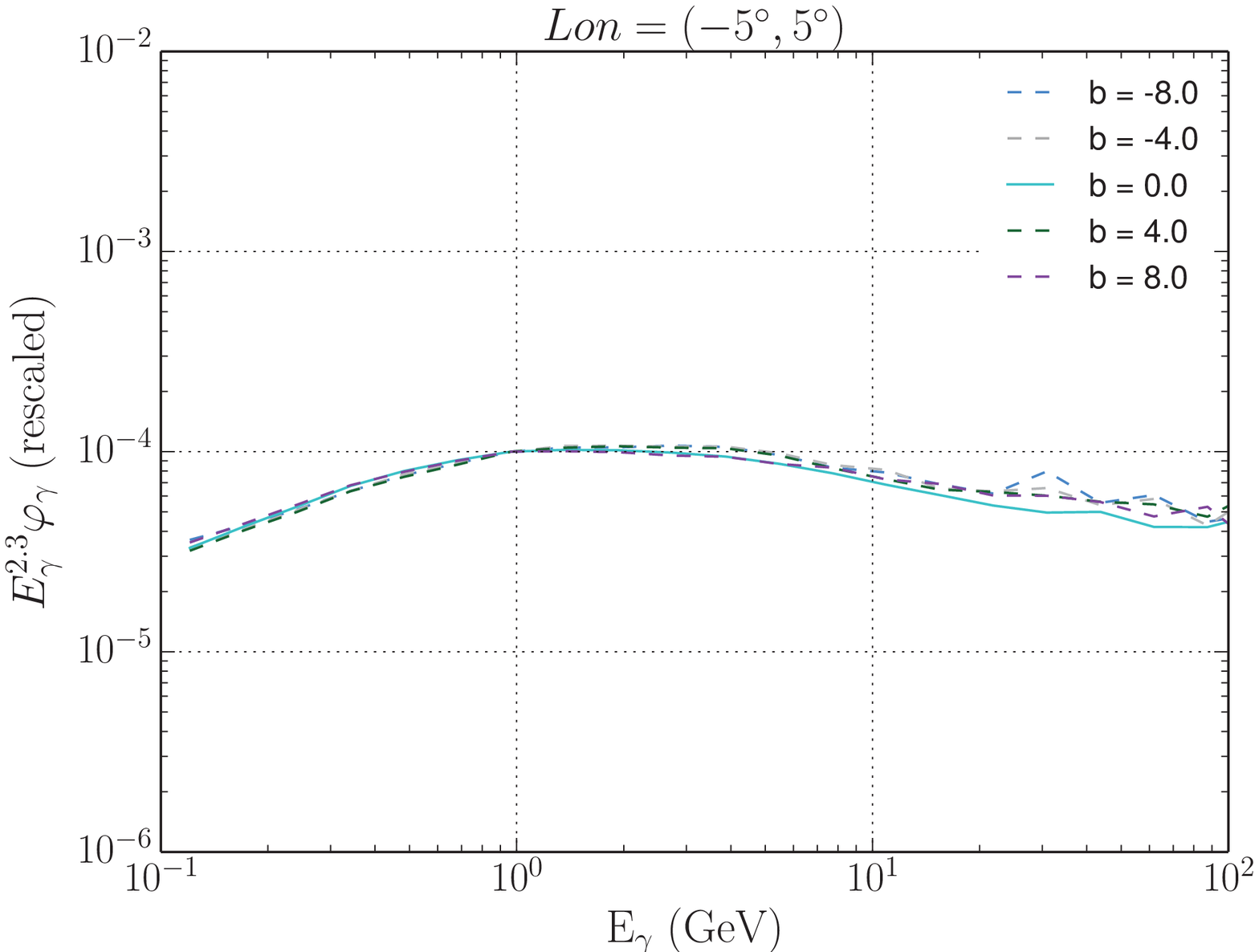}
\caption{
Latitude profiles of Galactic diffuse gamma-ray emission. The gamma-ray flux $\varphi_\gamma$ is obtained from the
total \Fermi-LAT data by subtracting the point sources and the isotropic components in the Sample model of \cite{acker17a}.
For comparison, the plotted values of $E_\gamma^{2.3}\varphi_\gamma$ are rescaled to $10^{-4}~{\rm
GeV^{1.3}~s^{-1}~cm^{-2}~sr^{-1}}$ at $E_\gamma = 1$~GeV. Upper panel: diffuse emission without PSF correction. Lower panel:
diffuse emission corrected for the PSF smearing of the data. The modification seen in the lower panel at low energies is
explained by the worsening angular resolution of the \Fermi-LAT at energies below 1 GeV (see text for details).}
\label{fig:profile}
\end{figure}

In Figure~\ref{fig:profile} we present the rescaled latitude profile of the gamma-ray data above and below the GC after
subtracting the contribution of point sources and isotropic emission. In this plot, we use the same data sample as in
\citet{acker17a}, i.e., 6.5 years of Pass 8 UltraCleanVeto Fermi-LAT data. The point sources are modeled using the 3FGL
catalog  \citep[see][]{acero15} with an overall rescaling factor determined in the Sample model of \citet{acker17a}. The
isotropic emission model is also taken from the Sample model of \citet{acker17a}. The upper panel shows the latitude profile
of the gamma-ray data after subtraction of emission from point sources and isotropic emission. Although there is an apparent
variation in the spectra at low energies, one has to take into account the increase of the point spread function (PSF) in
this case. The 68\% containment is worse than $1^\circ$ at $E_\gamma\lesssim700$~MeV, reaching about $6^\circ$ at
$E_\gamma\approx100$~MeV, which is larger than the width of the latitude stripes in the plot. Since the gamma-ray emission
is the brightest within $\approx 2^\circ$ from the Galactic Plane, the smearing of the data due to angular resolution
results in a depletion of the observed number of photons in the plane (relative to the expected flux with perfect angular
resolution). By the same token, the photons which leak from the plane will contaminate the flux at low energies at higher
latitudes. In order to correct for these effects, we calculate the correction coefficients for each energy bin and each
latitude stripe, by taking the ratio of the counts in Galactic diffuse gamma-ray emission model before and after the PSF
convolution of the model \citep[we use the Sample model of][]{acker17a}. The data multiplied by the correction coefficients
is shown in the lower panel of Figure~\ref{fig:profile}.

One can see that the spectra of the Galactic diffuse gamma-ray emission at different latitudes are consistent with each
other at low energies after the PSF correction. Several groups have studied the gamma-ray emission correlated with the gas
distribution in the inner Galaxy \citep{acker12, cal15, 2015PhRvD..91h3012G, acero16,ajello16,Daylan16}. Although at high
energies some differences are observed in the emissivity in the inner Galaxy compared to the average emissivity
\citep{2015PhRvD..91h3012G, acero16}, there is no appreciable hardening of the spectrum at lower energies, which is
consistent with the lower panel of Figure~\ref{fig:profile}.
This implies that available Fermi data do not allow us to draw reliable conclusions on possible modulation of CR proton
spectrum. We anticipate that data with a substantially higher resolution will be available with the next low-energy
gamma-ray missions, such as ASTROGAM \citep[see][]{astrogam} or AMEGO \citep[e.g.,][]{hart18}.


The other problem  of interpretation of diffuse gamma-ray flux from the CMZ is that it is model-dependent. The
above-mentioned GeV excess was determined  by subtracting (from the total observed flux) contributions of gamma-ray sources
and  known components of gamma-ray emission generated by CRs, which were calculated from models of CR propagation in the
Galaxy \citep[e.g., GALPROP, see][]{GALPROP}. After the  subtraction, a relatively small (but statistically significant)
component of emission is seen in the GC region, with the energy flux of $\sim 2\times10^{-7}$~GeV~s$^{-1}$~cm$^{-2}$ at
$E_\gamma\approx 2$~GeV, which stretches to $10^\circ$ or even $20^\circ$ radius from the GC \citep[see][]{hooper13,cal15}.
\citet{cal15} and \citet{acker17a} have pointed out that excesses of similar magnitude as in the GC are observed in other
locations along the Galactic Plane. As a result, the DM interpretation cannot be robustly confirmed.

\section{Self-consistent model of CR penetration into dense molecular clouds}
\label{self}

Attempts to analyse variations of CR spectra inside molecular clouds were undertaken long before the excess discovery.
Several mechanisms may cause these variations, such as particle acceleration inside and near molecular clouds
\citep{morf1,morf2,dog87,dog90}, or modulation of the CR flux by MHD turbulence excited in diffuse envelopes
\citep{skill76,cesar78, zweib11,gabici15,schlick16,dog15,ivlev18,phan}.

\citet{skill76} predicted from a qualitative analysis of the problem a depletion of CR density inside the clouds. They
concluded that  Alfven waves, generated by streaming CRs outside the clouds, suppress penetration of CRs with energies below
a few hundred of MeV. They mentioned also that GeV CRs could also be excluded, especially if their density near the clouds
is increased. This statement was confirmed by our analytical and numerical calculations (Paper I). Our investigations showed
also that the flux penetrating into {\it very dense} clouds (where CRs are fully absorbed) has a universal energy dependence
-- it is exclusively determined by the densities of ionized and neutral components of the cloud envelopes. Furthermore, in
Paper I we provided a detailed analysis of the CR-induced MHD turbulence near the clouds and determined the region where the
turbulence is excited. We showed that the turbulence leading to the universal flux can be generated either due to CR
absorption in a vicinity of dense cloud, or due to energy losses by CRs in the outer part of the envelope. In both cases,
the turbulence makes the spectrum of low-energy CRs harder and independent from that in the interstellar medium.

Let us summarize the principal results of Paper I, which are important for the analysis of gamma-ray emission presented in
Sections~\ref{gamma_CMZ} and \ref{gamma_local}. If the kinetic energy $E$ of CRs penetrating into a cloud exceeds a certain
{\it excitation threshold} $E_{\rm ex}$, their flux is not modulated, i.e., the propagation of such CRs through the cloud
envelope occurs in a free-streaming regime. For $E\lesssim E_{\rm ex}$, the CR propagation is diffusive -- due to efficient
pitch-angle scattering on the self-excited MHD turbulence, the mean free path of CRs is smaller than the relevant spatial
scale, and their pitch-angle distribution becomes quasi-isotropic. The resulting CR flux is usually strongly modulated.

The threshold $E_{\rm ex}$ for CR protons can be derived from the balance of the growth rate of MHD waves (calculated in the
free-streaming regime of CR propagation) and the damping rate due to ion-neutral collisions (Paper I):
\begin{equation}\label{threshold}
\frac{\tilde E_{\rm ex}+2}{\tilde E_{\rm ex}+1}\tilde E_{\rm ex}\tilde j_{\rm IS}(E_{\rm ex})
=2\epsilon \nu\,,
\end{equation}
where $\tilde E=E/m_{\rm p}c^2$ is the dimensionless kinetic energy normalized by the proton rest-mass energy, and $\tilde
j_{\rm IS}(E)=j_{\rm IS}(E)/j_*$ is the dimensionless energy spectrum of interstellar CRs, normalized by its characteristic
value $j_*=j_{\rm IS}(E=m_{\rm p}c^2)$. The threshold is determined by two dimensionless numbers, the damping rate $\nu$ and
small parameter $\epsilon=v_{\rm A}/c$ (ratio of the Alfven velocity to the speed of light). The scaling dependence of $\nu$
and $\epsilon$ on the physical parameters of the envelope is given by the following general expressions:\footnote{We
note that in Paper I, Equation~(23), there is a misprint in the normalization factor of $j_*$.}
\begin{align}
\nu& = 8.7 \left(\frac{m_{\rm g}/m_{\rm p}}{2.3}\right)
    \left(\frac{j_* m_{\rm p} c^2}{0.57~\mbox{cm}^{-2}\mbox{s}^{-1}\mbox{sr}^{-1}}\right)^{-1} \label{scale_nu}\\
	&\times\left(\frac{n_{\rm i}/n_{\rm g}}{3\times 10^{-4}}\right)\left(\frac{n_{\rm g}}{100~\mbox{cm}^{-3}}\right)^2
    \left(\frac{B}{0.1~\mbox{mG}}\right)^{-1},\nonumber \\
\epsilon&= 1.2\times10^{-3} \left(\frac{m_{\rm i}/m_{\rm p}}{12}\right)^{-1/2}\label{scale_epsilon}\\
	&\times \left(\frac{n_{\rm i}/n_{\rm g}}{3\times 10^{-4}}\right)^{-1/2}\left(\frac{n_{\rm g}}{100~\mbox{cm}^{-3}}\right)^{-1/2}
    \left(\frac{B}{0.1~\mbox{mG}}\right),\nonumber
\end{align}
where $m_{\rm g}$ is the average mass of gas particles, $n_{\rm g}$ and $n_{\rm i}$ is the density of gas particles and
ions, respectively, and $B$ is the magnetic field strength.

The modulated CR flux (per unit area and unit energy interval), propagating through the envelope and entering the dense
interior of the cloud, is given by
\begin{equation}
\label{flux}
S(E)=\frac{v_{\rm A}N_{\rm IS}(E)}{1-e^{-\eta_0(E)}}\,,
\end{equation}
where $N_{\rm IS}=j_{\rm IS}/(4\pi v)$ is the differential density of interstellar CRs and $\eta_0$ is a measure of the
relative importance of diffusion and advection in the modulated flux: For $v_{\rm A}/v\lesssim\eta_0\lesssim1$ the flux is
dominated by diffusion, while for large $\eta_0$ the interstellar CR density is advected with the Alfven velocity $v_{\rm
A}$. As we showed in Paper I, the value of $\eta_0$ in Equation~(\ref{flux}) can be approximately presented as
\begin{equation}\label{zeta_0}
\eta_0(E)\approx\sqrt{\tilde E(\tilde E+2)}\;\frac{\tilde j_{\rm IS}(E)}{2\nu}\,,
\end{equation}
valid as long as $\eta_0\lesssim1$ (otherwise, its value is unimportant for calculating the flux).

All results presented in sections below are obtained for the following energy spectrum of interstellar CR:
\begin{equation}\label{CMZ}
j_{\rm IS}(E)=j_*\left(\frac2{\tilde E+1}\right)^{2.8} \,,
\end{equation}
where $j_* =0.29~{\rm cm}^{-2}{\rm s}^{-1}{\rm sr}^{-1}{\rm GeV}^{-1}$ corresponds to the local spectrum of CR protons.
Equation~(\ref{CMZ}) was derived from the local gamma-ray emissivity using the pp cross section from \citet{ppcross_1} and
\cite{ppcross_2}, with the emissivity estimated by \citet{casa15} from high-latitude Galactic gamma-ray emission \citep[see
also][]{str15}.

\subsection{Effect of finite cloud size and heavier CR species}
\label{finite}

First, we need to take into account that dense clouds have finite sizes, and therefore most of the penetrating CRs cross the
clouds without absorption.  The flux velocity $u$ of these CRs, which are assume to stream freely through the cloud, is
not equal to their physical velocity $v$, but to a certain smaller value determined by the cloud size. One can easily
evaluate $u$ for relativistic protons, where the pion production losses dominate (for lower energies, also ionization losses
can be added, see Appendix~\ref{apx:CRn}): The flux of CRs entering a cloud from either size (along the  local magnetic
field lines) is approximately equal to $c(N_{\rm in}-N_{\rm out})$, where $N_{\rm in}$ is the local density of incident
CRs. The density of CRs that crossed the cloud is attenuated due to pp collisions, $N_{\rm out}\approx N_{\rm
in}\exp(-2\sigma_{\rm pp}\mathcal{N}_{\rm H_2})$; it is determined by the product of the pion production cross section,
$\sigma_{\rm pp}$, and the effective column density of the cloud, $\mathcal{N}_{\rm H_2}$. Then the flux velocity  at
the cloud edge is
\begin{equation}\label{u}
u \approx \frac{N_{\rm in} - N_{\rm out}}{N_{\rm in} + N_{\rm out}}\:c\approx\frac12\kappa c,
\end{equation}
where we took into account that the product $2\sigma_{\rm pp}\mathcal{N}_{\rm H_2}\equiv\kappa$ is typically a small number.
Since the growth rate of MHD waves excited by free-streaming CRs is proportional to their flux, we immediately conclude that
the left-hand side (lhs) of Equation~(\ref{threshold}) should be multiplied by a small factor of $\frac12\kappa$. Hence, the
excitation threshold $E_{\rm ex}$ is reduced for a finite cloud size  (i.e., $E_{\rm ex}$ is a certain increasing
function of $\kappa$, its form is determined by the interstellar CR spectrum).

Next, we consider the effect of finite cloud size on the flux $S(E)$ in the diffusive regime, $E\lesssim E_{\rm ex}$, where
CRs are self-modulated while propagate through an envelope toward a dense cloud. For a perfectly absorbing wall studied in
Paper~I (see Figure~1 of of that paper), $S(E)$ is described by Equation~(\ref{flux}); the local CR density $N(z)$ gradually
decreases as particles approach the wall located at $z=0$, and therefore the local flux velocity $S/N$ increases. For a
finite cloud size, the latter cannot exceed the velocity $u$ in the free-streaming regime, given by Equation~(\ref{u}).
Therefore, the problem can be analyzed exactly in the same manner as in Paper~I, with the only difference that the boundary
condition at the cloud border is given by
\begin{equation}\label{bond1}
N(0)=S/u.
\end{equation}
This implies that the CR density at $z=0$ is higher than that in the absorbing-wall case (by a factor of $\sim c/u$ for
relativistic CRs).\footnote{The physical reason for this enhancement is that particles in a finite-size cloud have to be
trapped for their attenuation time.} The resulting modulated CR flux, corrected for the ``inner'' boundary
condition~(\ref{bond1}), is
\begin{equation}
\label{flux1}
S(E)=\frac{v_{\rm A}N_{\rm IS}(E)}{1-(1-v_A/u)e^{-\eta_0(E)}}\,.
\end{equation}
As in Equation~(\ref{flux}), $S(E)$  tends to the advection asymptote $v_{\rm A}N_{\rm IS}$ for $\eta_0\ga 1$, whereas
for $\eta_0 \lesssim1$ the flux has the universal diffusion-dominated form $S_{\rm DD}(E)$ (Equation~(42) of Paper I)
limited by the maximum value of $uN_{\rm IS}$.

\begin{figure}[h]
\centering
\includegraphics[width=\columnwidth]{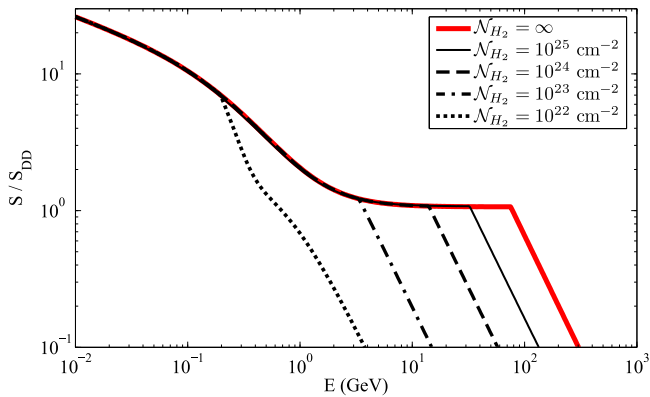}
\caption{ The ratio of the total CR flux $S(E)$ to its universal (diffusion-dominated) component $S_{\rm DD}(E)$. Above
the kink, $E>E_{\rm ex}$, the curves represent the free-streaming flux $uN_{\rm IS}$ for different values of
$\mathcal{N}_{\rm H_2}$. The modulated flux below the kink is described by Equation~(\ref{flux1}). The shown results are
for $\nu=0.39$ and for the interstellar (non-modulated) spectrum~(\ref{CMZ}).} \label{SDD}
\end{figure}

In Figure~\ref{SDD} we show the ratio of the total modulated flux (penetrating into the cloud), $S(E)$, to its universal
component, $S_{\rm DD}(E)$. The solid line represents the case of a perfectly absorbing wall, the other lines show cases of
finite-size clouds. One can see that for sufficiently large values of $\mathcal{N}_{\rm H_2}$ the flux is universal in a
rather broad energy range (below $E_{\rm ex}$, where the curves are practically horizontal), as expected from the
excitation-damping balance. For smaller $\mathcal{N}_{\rm H_2}$ the balance is violated and the flux approaches the
advection asymptote.

The excitation threshold $E_{\rm ex}$ for finite values of $\mathcal{N}_{\rm H_2}$ is derived from the modified
Equation~(\ref{threshold}). For sufficiently large values of $\mathcal{N}_{\rm H_2}$, the threshold is in the relativistic
energy range, where $j_{\rm IS}(E)\propto E^{-2.8}$. In this case, $E_{\rm ex}$ can be determined from a simple scaling
dependence,
   \begin{eqnarray}
    \left(\frac{E_{\rm ex}}{5\mbox{ GeV}}\right)^{1.8} && \approx \left(\frac{m_{\rm g}/m_{\rm p}}{2.3}\right)^{-1}
    \left(\frac{m_{\rm i}/m_{\rm p}}{12}\right)^{1/2} \label{scale_Ex} \\
    && \times \left(\frac{j_{5} m_{\rm p} c^2}{10^{-2}~\mbox{cm}^{-2}\mbox{s}^{-1}\mbox{sr}^{-1}}\right) \nonumber\\
    &&\times\left(\frac{n_{\rm i}/n_{\rm g}}{3\times 10^{-4}}\right)^{-1/2}
		\left(\frac{n_{\rm g}}{10~\mbox{cm}^{-3}}\right)^{-3/2} \nonumber\\
    &&\times \left(\frac{\mathcal{N}_{\rm H_2}}{2\times10^{23}~\mbox{cm}^2}\right) \,, \nonumber
   \end{eqnarray}
Instead of the rest-mass energy, we chose $E=5$~GeV as a convenient energy scale for this case, so that the CR spectrum is
normalized by $j_5=j_{\rm IS}(E=5~{\rm GeV})$. We point out that the effect on the gamma-ray emission is only observable if
$E_{\rm ex}$ exceeds a certain value (which, obviously, cannot be smaller than the pion production threshold of
$\approx280$~MeV). In practice, gamma-ray emission should be modified at $E_\gamma\gtrsim100$~MeV, which roughly corresponds
to $E_{\rm ex}\gtrsim1$~GeV. In Table~\ref{concl:table:NHEex} we present the threshold values of the effective column
density, $\mathcal{N}_{H_2}^{\rm tr}$, needed to reach the excitation threshold of $E_{\rm ex}\approx1$~GeV for given values
of the gas density and the magnetic field in the envelope.

\begin{table}[h]
\caption{ The threshold column density $\mathcal{N}_{\rm H_2}^{\rm tr}$ (in units of $10^{23}$~cm$^{-2}$) for $E_{\rm ex}
\approx 1$~GeV, assuming  the interstellar spectrum~(\ref{CMZ}) and varying gas density $n_{\rm g}$ (in units of
cm$^{-3}$) and magnetic field $B$ (in units of $\mu$G).} \centering
\begin{tabular}{|l|c|c|c|c|c|}
\hline
&  $n_{\rm g} = 1$ & $n_{\rm g} = 3$ & $n_{\rm g} = 10$ & $n_{\rm g} = 30$ & $n_{\rm g} = 100$ \\
\hline
$B = 1$& 0.06 & 0.07 & 0.34 & 1.65 & 11 \\
\hline
$B = 3$& 0.15 & 0.14 & 0.36 & 1.65 & 11\\
\hline
$B = 10$& 0.44 & 0.32 & 0.46 & 1.7 & 11 \\
\hline
$B = 30$& 1.3 & 0.9 & 0.75 & 1.9 & 11 \\
\hline
\end{tabular}
\label{concl:table:NHEex}
\end{table}

Finally, we notice that about 10\% of interstellar CRs are helium nuclei and heavier ions. The rate of wave excitation by
nuclei of charge $Z$ is given by Equation~(\ref{eq:apx1:gamma}) of Appendix~\ref{apx:CRn}. The absorption of heavier CR
nuclei in a dense cloud is dominated by their spallation, with the cross section
\begin{equation}
\sigma_Z\approx 1.5\sigma_{\rm pp}A^{0.7},
\label{ssp}
\end{equation}
where $A(Z)$ is the atomic mass number \citep[see e.g.][]{mann94}. This makes the corresponding flux velocity $u_Z$ higher
than that of protons. A detailed analysis presented in Appendix~\ref{apx:CRn} shows that the resulting effect of heavier
species on the wave excitation can be substantial.

\section{CR modulation and gamma-ray emission in the CMZ region}
\label{gamma_CMZ}

The density of CRs in the CMZ region is not known, although estimates  suggest that it can be a factor of a few larger
than the local interstellar spectrum \citep[see][]{yang2, acero16}. We assume that the proton spectrum near CMZ is {\it five
times} the local spectrum given by Equation~(\ref{CMZ}), and use the following reasonable parameters of the CMZ region:
the gas density in the envelope $n_{\rm g}=10$~cm$^{-3}$, the magnetic field strength $B=10~\mu$G, and the effective CMZ
column density $\mathcal{N}_{\rm H_2}=10^{23}$~cm$^{-2}$.

\begin{figure}[h]
\centering
\includegraphics[width=\columnwidth]{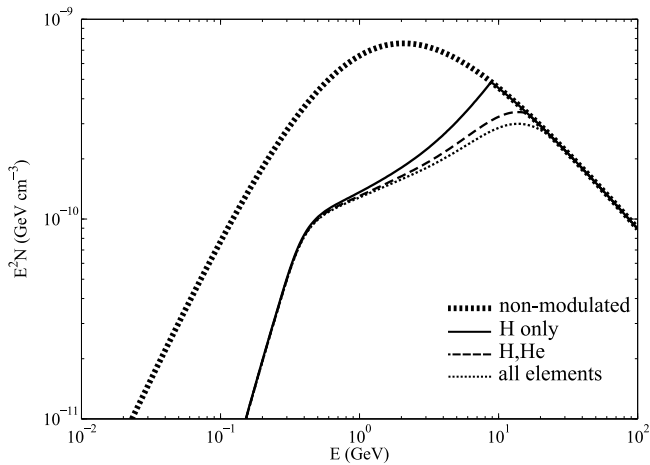}
\caption{The modulated energy spectrum of CR protons in the CMZ, calculated for the interstellar spectrum~(\ref{CMZ}) (thick
dotted line). The solid lines represent the case where no effects of heavier CRs are included, the other curves are for
additional 10\% of He nuclei (dashed lines) and for all elements (dotted line). The hydrogen density in the envelope is
$10$~cm$^{-3}$, the magnetic field strength is $10~\mu$G, the effective gas column density of the CMZ is
$10^{23}$~cm$^{-2}$.} \label{proton}
\end{figure}

By applying the self-consistent model developed in Paper I, and including the corrections discussed in Section~\ref{finite},
we calculate the CR proton spectrum in the CMZ. The solid line in Figure~\ref{proton} shows the results for the case where
effects of heavier CR species is neglected, Equation~(\ref{flux1}). The effect of heavier CRs is also illustrated: one can
see that the peak at $E_{\rm ex}$ becomes smoother, and the position of the excitation threshold shifts to higher energies.
The curve representing all elements includes contributions of nuclei whose excitation amplitude $\chi_Z$ exceeds the value
of 0.1 (see Appendix~\ref{apx:CRn} and Table~\ref{apx1:table:CRs} therein).

\begin{figure}
\centering
\includegraphics[width=\columnwidth]{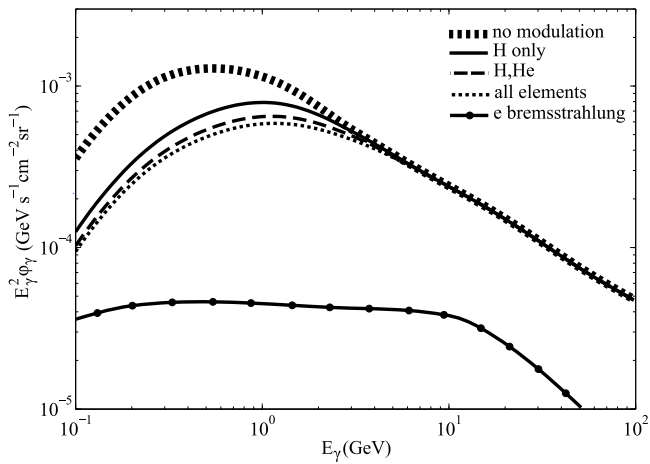}
\caption{
The expected CMZ gamma-ray fluxes, calculated for the CR proton spectra presented in Figure~\ref{proton}. The gamma-ray flux
for the interstellar (non-modulated) spectrum~(\ref{CMZ}) is shown by the thick dotted line. The contribution of the CR
electron bremsstrahlung is also plotted for comparison (solid line with dots).} \label{gamma}
\end{figure}

Using the derived spectra of CRs, the expected gamma-ray flux from the CMZ can be calculated as
\begin{eqnarray}\label{eq:eq_gammarays}
\varphi_\gamma(E_\gamma) =&&\mathcal{N}_{\rm H_2}c\nonumber\\
&& \times\left[\sum_Z\int dE_Z~N_Z(E_Z) \left(\frac{d\sigma(E_Z,E_\gamma)}{dE_\gamma}\right)_{\rm pp}\right.\nonumber\\
&&\left.+\int dE_e ~N_e(E_e)  \left(\frac{d\sigma(E_e,E_\gamma)}{dE_\gamma}\right)_{\rm br} \right] \,,
\end{eqnarray}
where the differential cross section of the photon production due to the $\pi^0$ decay, $(d\sigma/dE_\gamma)_{\rm pp}$, is
taken from \citet{ppcross_1} and \citet{ppcross_2}, the bremsstrahlung cross sections, $(d\sigma/dE_\gamma)_{\rm br}$, are
from \citet{blu70} for the electron-nuclear interactions and from \citet{haug} for the electron-electron interactions.
Figure~\ref{gamma} shows the calculated $\varphi_\gamma(E_\gamma)$. We see that the gamma-ray flux due to the modulated CRs
differs significantly from the results derived in the framework of the standard approach (no modulation). The peak of
gamma-ray emission shifts to higher energies, and its position depends on the gas density in the envelope and the density of
interstellar CRs.

The presented results suggest that the procedure used for calculating the CR contribution to the total gamma-ray flux should
be revised for the CMZ region. We point out, however, that the expected effect, shown in Figure~\ref{gamma}, cannot be
reliably confirmed or discarded with the available data, due to their poor resolution at $E_\gamma\lesssim1$~GeV (see
Section~\ref{fermi}).

\begin{figure*}
\centering
\includegraphics[width=.9\textwidth]{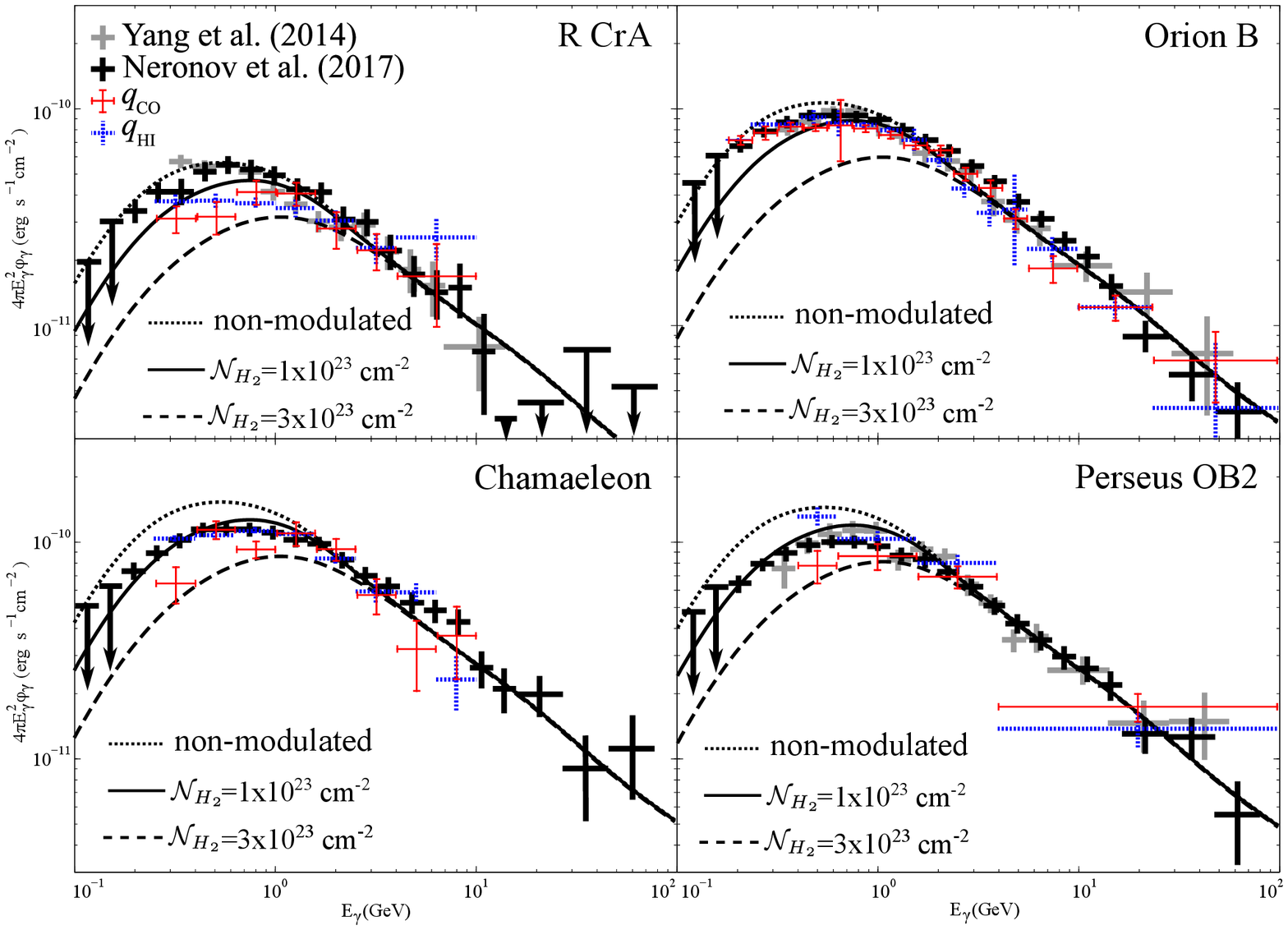}
\caption{The measure gamma-ray fluxes from the clouds R CrA, Orion B, Chamaeleon, and Perseus OB2, by \citet{yang1} (thick
gray crosses) and \citet{ner17} (thick black crosses). The (normalized) fluxes estimated by the \Fermi team (see references
in the text) from the emissivity of atomic hydrogen ($q_{\rm HI}$) and molecular gas ($q_{\rm CO}$) are depicted by the thin
blue and red crosses, respectively. The calculated flux due to the non-modulated CR spectrum~(\ref{CMZ}) is shown by the
dotted line; the fluxes modulated by the effective column densities of $10^{23}$~cm$^{-2}$ and $3\times10^{23}$~cm$^{-2}$
are represented by the solid and dashed lines, respectively.} \label{local}
\end{figure*}

\section{Gamma rays from local molecular clouds}
\label{gamma_local}

The effect of CR modulation discussed above can, in principle, be observed in local molecular clouds. These clouds are
located at distances of a few hundreds of pc from the Earth and, therefore, are spatially resolved with the \Fermi-LAT
telescope. If present, the modulation effect should manifest itself with a decreased gamma-ray emissivity at lower energies
inside the cloud, as compared to the interstellar emissivity.

Recent publications on gamma-ray emission from local clouds \citep[see][]{yang1,ner17,remy17} presented rather controversial
conclusions about the density of CRs inside the clouds. While \citet{ner17} and \citet{remy17} did not find any difference
between the local CR spectrum near the Earth and that inside the clouds below 18~GeV, \citet{yang1} claimed a significant
modulation of CR density at energies below 10~GeV in several clouds (as identified from the Pamela data).

In Figure~\ref{local} we plot the gamma-ray fluxes $\varphi_\gamma(E_\gamma)$ obtained by \citet{yang1} for the clouds
R~CrA, Orion, and Perseus OB2 (thick gray crosses), and compare these with the unmodulated flux calculated for the CR
spectrum defined in Equation (\ref{CMZ}) (dotted lines). We also show the data from \citet{ner17} for these clouds and for
the cloud Chamaeleon (thick black crosses), normalized to the unmodulated flux at 10~GeV. One can see that the data for
clouds R~CrA and Orion do not reveal significant modulation, while for Perseus and Chamaeleon the effect is visible.

For several clouds, measurements of atomic hydrogen emissivity $q_{\rm HI}(E_\gamma)$ and molecular gas emissivity  $q_{\rm
CO}(E_\gamma)$ (traced by CO emission lines) are available \citep[e.g.][]{acker12b, acker13, remy17}. Since the gas in the
outer diffuse regions of the clouds is atomic, while the molecular gas is concentrated in the dense cores, the emissivity
$q_{\rm HI}$ characterizes the CR spectrum in the cloud envelopes, and $q_{\rm CO}$ is a measure of the CR spectrum inside
the clouds. Comparison of $q_{\rm CO}(E_\gamma)$ with $q_{\rm HI}(E_\gamma)$ gives us a direct estimate of the modulation
inside a cloud, independent of possible variations of the interstellar spectrum across the Galaxy. In Figure~\ref{local} we
presented gamma-ray fluxes for clouds R~CrA \citep{acker12b}, Orion \citep{remy17}, Chamaeleon \citep{acker13}, and Perseus
\citep{remy17}, deduced from $q_{\rm HI}$ (thin blue crosses) and $q_{\rm CO}$ (thin red crosses). Again, we see that R~CrA
and Orion do not show significant modulation (which is also the case for most of the clouds analyzed in these papers), while
for Chamaeleon and Perseus a marginal modulation effect is present at lower energies.

We recall that, according to our model, the excitation threshold is a function of the {\it effective column density}
$\mathcal{N}_{\rm H_2}$, measured along the magnetic field line. In Figure~\ref{local} we illustrate the expected modulation
effect for $\mathcal{N}_{\rm H_2}= 10^{23}$~cm$^{-2}$ (solid line) and $\mathcal{N}_{\rm H_2}= 3\times10^{23}$~cm$^{-2}$
(dashed line). For example, the substantial difference between the modulated and non-modulated fluxes for Chamaeleon and
Perseus can be explained if the effective column density is of the order of $10^{23}$~cm$^{-2}$. We point out that
$\mathcal{N}_{\rm H_2}$ may significantly exceed the values deduced from the line-of-sight measurements, since the magnetic
field lines in molecular clouds can be twisted \citep[see e.g.][]{dog87,padovani13}, or the clouds can be elongated along
the field lines, and the magnetic field can thread multiple molecular clouds \citep{planck15}. However, all these effects
require separated analysis \citep[see, e.g.,][]{ivlev18b}, which is beyond the scope of our paper.

Thus, there are evidences of the modulation effect in local molecular clouds. Several observations of gamma-ray emission
from the clouds indicate the depletion of CR density at lower energy, as suggested in Section~\ref{self}. We hope that
future observations will provide information for more reliable analysis.

\section{ Conclusions}
\label{discussion}

 We investigated the effect of finite-size molecular clouds on the energy spectrum of the penetrating CRs, and on the
resulting gamma-ray emission. The conclusions of the paper can be summarized as follows:
\begin{itemize}
\item The energy dependence of self-modulated CR flux penetrating into a cloud is exactly the same as in Paper~I: For
    sufficiently high effective column densities, the flux has the universal form $S_{\rm DD}(E)$, independent of the
    interstellar CR spectrum. For lower column densities, the nonlinear wave cascade becomes important and the flux
    velocity approaches the Alfven velocity $v_{\rm A}$;
\item The probability of relativistic CRs to be absorbed in a finite-size molecular cloud, characterized by the column
    density $\mathcal{N}_{\rm H_2}$, is typically small and equal to $\sigma_{\rm pp} \mathcal{N}_{\rm H_2}\ll1$, where
    $\sigma_{\rm pp}$ is the effective cross section of the pion production. The resulting velocity of the CR flux
    entering the cloud is $u\approx \sigma_{\rm pp}\mathcal{N}_{\rm H_2}v\ll v$. This leads to a significant
    reduction of the increment of MHD wave excitation, and therefore makes the value of the excitation threshold energy
    $E_{\rm ex}$ (below which the CR flux is modulated) smaller than in the case of a perfectly absorbing wall (Paper
    I);
\item The density of modulated CRs inside a finite-size central core, between the two turbulent regions self-generated
    in a diffuse envelope, is $N\sim S_{\rm DD}/u$. Typically, this is much larger than the CR density in a perfectly
    absorbing core, $N\sim S_{\rm DD}/v$, though in both cases these values are still significantly smaller than the
    interstellar density $N_{\rm IS}$;
\item For typical parameters of the CMZ, the self-modulation of penetrating CRs is expected in the energy range below
    $E\sim 10$~GeV. The resulting modulation of gamma-ray flux occurs at energies below $E_\gamma\approx2$~GeV;
\item The phenomenon of self-modulation should be taken into account when evaluating the GeV excess in the total
    gamma-ray emission produced in the GC. We showed that the emission due to self-modulated CRs may substantially
    deviate from predictions of the standard model of CR propagation in the Galaxy \citep[e.g., GALPROP,][]{GALPROP}.
    This implies that both the value and the form of the ``GeV excess'' -- obtained after subtracting different
    components (including that generated by CRs) from the observed emission -- may be quite different from what is
    currently believed;
\item The worsening spatial and energy resolution of the \Fermi-LAT telescope at energies below $E_\gamma\sim1$~GeV does
    not allow us to draw reliable conclusions about the intensity and spectrum of gamma-ray emission from the CMZ
    region. Furthermore, disentangling of the emission due to the interaction of CR protons and CMZ gas from other
    gamma-ray sources in the GC, including Sgr A$^*$, is a challenging task with the available resolution. We note that
    the GC environment is unique, and it is currently difficult to definitively determine the interplay between CR
    modulation and several additional complexities in the GC (see Section~\ref{fermi});
\item The effect of CR self-modulation in the gamma-ray emission can possibly be detected by observing nearby molecular
    clouds, located at distances of a few hundred pc from Earth and, therefore, resolvable with the \Fermi-LAT
    telescope. While there are indications of this effect for several nearby clouds, observable depletion of the
    emission is predicted to occur only for $\mathcal{N}_{\rm H_2}$ as high as $\sim10^{23}$~cm$^{-2}$. This is much
    larger than typical line-of-sight column densities of less than $\sim10^{22}$~cm$^{-2}$, as derived from CO
    observations. On the other hand, the effective column density $\mathcal{N}_{\rm H_2}$ is likely to significantly
    exceed the values deduced from the line-of-sight measurements and, hence, to reach the values needed for observable
    depletion of the emission.
\end{itemize}

\section*{Acknowledgments}

The authors are grateful to Wim de Boer for fruitful discussions, and to Isabelle Grenier, Tsunefumi Mizuno, and Luigi
Tibaldo for helpful comments on measurements of gamma-ray emission in molecular clouds.  VAD and DOC are supported by the
grant RFBR 18-02-00075. DOC is supported in parts by foundation for the advancement of theoretical physics ``BASIS''. KSC is
supported by the GRF Grant under HKU 17310916. The authors also would like to thank the anonymous referee for constructive
comments.

\software{
GALPROP \citep[\url{http://galprop.stanford.edu/},][]{GALPROP},
HEALPix \citep[\url{http://healpix.jpl.nasa.gov/},][]{healpix},
matplotlib \citep{matplotlib}
}.

\appendix

\section{A. Modulation of CR flux with multiple nuclei}
 \label{apx:CRn}

The spectra of CRs include different nuclei. Although the proton abundance is about 90\%, the effect of heavier nuclei on
the excitation of MHD-waves should also be taken into account. The rate of the wave excitation by CR species with the charge
$Ze$ is \citep[see, e.g.][]{kuls69}
\begin{equation}
\gamma_Z = \pi^2 \frac{Ze v_{\rm A}}{cB}p_Zv_Z \left(S_Z - v_{\rm A}N_Z\right) \equiv \pi^2 \frac{Z^2 e^2 v_{\rm A}}{c^2B}
Rv_Z\left(S_Z - v_{\rm A}N_Z\right) \,.
\label{eq:apx1:gamma}
\end{equation}
Here, $S_Z(E)$ is the flux of nuclei $Z$ in the diffusive regime of CR propagation, and $R=p_Zc/Ze$ is the magnetic
rigidity, which is proportional to the momentum $p_Z$ of a nucleus. The use of the rigidity in
Equation~(\ref{eq:apx1:gamma}) is convenient, since it is related to the resonant wave number via $k = B/R$ (independent of
$Z$). Furthermore, this allows us to utilize CR databases (presented in units of $R$), assuming that the spectra of protons
and heavier nuclei have the same dependence on $R$.

The modulated CR flux $S_Z$, given by Equation~(\ref{flux1}), depends on two parameters, $\eta_{0,Z}(E_Z)$ and $N_Z(E_Z)$.
By using the definition of $\eta_0$ from Paper~I (see Equation~(27) of that paper) and expressing it through $R$, we obtain
that $\eta_{0,Z}/\eta_0 = {v_Z}/{v_1}$, where $\eta_0 \equiv \eta_{0,1}$ and $v_1$ correspond to protons and $v_Z$ is the
velocity of nucleus $Z$:
\begin{equation}
\frac{v_Z}{c} = \frac{Z\tilde R}{\sqrt{Z^2\tilde R^2 + A^2}} \,,
\end{equation}
where $\tilde R = Re/m_{\rm p}c^2$ and $A(Z)$ is the atomic mass number. Thus, we can write $\eta_{0,Z}(R) =
\xi_Z(R)\eta_0(R)$, where
\begin{equation}
\xi_Z(R) = \sqrt{\frac{\tilde R^2 + (A/Z)^2}{\tilde R^2 + 1}}
\end{equation}
is a numerical factor, varying between $A/Z$ for $\tilde R\ll1$ and $1$ for $\tilde R\gg1$.

The density $N_Z(E_Z)$ is related to the corresponding energy spectrum via $N_Z = j_Z/(4\pi v_Z)$, while the relation
between the spectra per unit energy and rigidity is $E_Zj_Z(E_Z) = Rj_Z(R)$ (in the ultra-relativistic case). This yields
\begin{equation}
\tilde{j}_Z(E_Z) = \frac{1}{Z} \frac{j_Z(R)}{j_1(R)} \tilde{j}_{\rm IS}(E)\,,
\end{equation}
where $\tilde j_{\rm IS}(E)=j_{\rm IS}(E)/j_*$ is the dimensionless interstellar spectrum and $E$ has to be expressed in
terms of $R$. We assume that the energy spectra for all CR species are given by Equation~(\ref{CMZ}), where $\tilde E$
should be replaced by the effective dimensionless energy $\sqrt{\tilde R^2 + 1} - 1$. The abundances of CR nuclei from the
CRDB database \citep{crdb} are summarized in Table~\ref{apx1:table:CRs},  where we list the species important for the wave
excitation, as explained below.

If we include contributions of all CR species into the process of MHD-wave excitation, the excitation-damping balance
at the outer bound of the diffusion zone (cf. rhs of Equation~(29) in Paper I) becomes
\begin{equation}
\tilde R\sum \limits_Z Z^2\tilde{j}_Z\frac{\delta_Z e^{-\xi_Z\eta_0}}{1- \delta_Z e^{-\xi_Z\eta_0}} = 2\nu\,,
\label{eq:apx1:balance}
\end{equation}
where $\delta_Z = 1 - v_{\rm A}/u_Z$ and the relation $\tilde R=\tilde k^{-1}$ is taken into account. The flux velocity
$u_Z$ of species $Z$ is determined by a combination of the catastrophic and continuous energy losses,
\begin{equation}
u_Z = \min\left\{v\mathcal{N}_{\rm H_2}\left(\sigma_Z+ b\frac{L_Z}{E_Z} \right),v\right\} \,,
\end{equation}
where $\sigma_Z$ is the total cross section of catastrophic (spallation) collisions, related to the pp cross section by
Equation~(\ref{ssp}), and $L_Z = -\dot E_Z/n_{\rm g}v_Z$ is the continuous (ionization) loss function;  the factor
$b\sim1$ is determined by the form of the local CR spectrum at the cloud edge. In the relativistic case, where the
catastrophic losses dominate, the flux velocities are
\begin{eqnarray}
&u_1/c = \frac{1}{2}\kappa & \mbox{(protons)}; \\
&u_Z/u_1 \approx 1.5A^{0.7} & \mbox{(heavier nuclei)},
\end{eqnarray}
where $\kappa \equiv 2\mathcal{N}_{\rm H_2}\sigma_{\rm pp}$. Equation~(\ref{eq:apx1:balance}) can be resolved for $\eta_0$ as
a function of rigidity, which allows us to calculate the CR spectra from Equation~(\ref{flux1}).  For very dense clouds, the
flux velocities are about the respective physical velocities and hence $\delta_Z\approx1$. Then, for small values of
$\eta_0$, one can expand the exponentials in the denominator of Equation~(\ref{eq:apx1:balance}) (and set them equal to
unity in the numerator). This leads to an analytical expression for $\eta_0(R)$, which is a straightforward generalization
of Equation~(\ref{zeta_0}) for multiple CR species (taking into account the above relation between $\tilde R$ and $\tilde
E$).

\begin{table}
\caption{Interstellar abundances of major CR species and their partial excitation amplitudes, $\chi_Z$.} \centering
\begin{tabular}{|c|c|c|c|c|}
\hline
Element & $Z$ & $A$ & $Rj_Z(R)$ at $R =20$~GV, & $\chi_Z$ \\
 &  & & (m$^2$~s~sr)$^{-1}$ &  \\
\hline
H & 1 & 1 & 78 & 1 \\
\hline
He & 2 & 4 & 15 & 1.5 \\
\hline
C & 6 & 12 & 0.49 & 0.32 \\
\hline
O & 8 & 16 & 0.48 & 0.5 \\
\hline
Mg & 12 & 24 & 0.08 & 0.17 \\
\hline
Si & 14 & 28 & 0.07 & 0.19 \\
\hline
Fe & 26 & 56 & 0.07 & 0.59 \\
\hline
\end{tabular}
\label{apx1:table:CRs}
\end{table}

Equation~(\ref{eq:apx1:balance}) with the condition $\eta_0 = 0$ yields the excitation threshold  $R_{\rm ex}$ expressed
in terms of the rigidity (which then can be converted into the energy excitation thresholds for each CR species). The
resulting equation for $R_{\rm ex}$ is
\begin{equation}\label{a10}
\tilde R_{\rm ex}\tilde{j}_{\rm IS}(\tilde R_{\rm ex})\frac{u_1}{c}\sum \limits_Z \delta_Z \chi_Z =2\epsilon\nu \,,
\end{equation}
where $\chi_Z$ is the partial ``excitation amplitude''. For protons $\chi_1 = 1$, while for heavier nuclei it is
\begin{equation}
\chi_Z \approx 1.5 Z A^{0.7} \frac{j_Z(R)}{j_1(R)} \,.
\end{equation}
The value of $\chi_Z$ characterizes the relative contribution of CR species $Z$ to the wave excitation,  in
Table~\ref{apx1:table:CRs} we list all species with $\chi_Z>0.1$. Again, for very dense clouds we have $u_1\to v_1$ and
$\delta_Z\approx1$, and then Equation~(\ref{a10}) leads to a simple generalization of Equation~(\ref{threshold}) for
multiple species.

\end{document}